# High-speed high-resolution optical correlation-domain reflectometry without using electrical spectrum analyzer

Guangtao Zhu,[a,1] Tomoya Miyamae,[a,1] Kohei Noda,[a,b] Heeyoung Lee,[c]

Kentaro Nakamura,[b] and Yosuke Mizuno[a,*]

[a]Faculty of Engineering, Yokohama National University,

79-5 Tokiwadai, Hodogaya-ku, Yokohama 240-8501, Japan

[b]Institute of Innovative Research, Tokyo Institute of Technology,

4259, Nagatsuta-cho, Midori-ku, Yokohama 226-8503, Japan

[c]College of Engineering, Shibaura Institute of Technology,

3-7-5 Toyosu, Koto-ku, Tokyo 135-8548, Japan

**mizuno-yosuke-rg@ynu.ac.jp*

[1]These authors contributed equally to this work.

**Abstract**

We propose a newly configured optical correlation-domain reflectometry (OCDR) without the use of an electrical spectrum analyzer (ESA) for high-speed distributed reflectivity measurement with an enhanced spatial resolution. First, the operation of ESA-free OCDR is analyzed by numerical simulation, the results of which show the potential of the configuration to perform distributed reflectivity measurement with an improved spatial resolution. Subsequently, the simulation results are verified by comparison experiments, where distributed reflectivity measurements along a 32-m-long single-mode fiber joint with multiple connectors are performed using both conventional (with averaging) and ESA-free OCDR configurations (without averaging) at a repetition rate of 100 Hz. The widths of the reflectivity peaks experimentally obtained by the conventional and ESA-free OCDR configurations are 22.9 cm and 7.0 cm, respectively. Finally, we experimentally evaluated the performance of the ESA-free OCDR at higher repetition rates, achieving 6.3-cm spatial resolution at a repetition rate of 1 kHz. This result indicates that the ESA-free OCDR can perform a distributed reflectivity measurement with an improved spatial resolution at a higher operation speed than ever.

## 1. Introduction

In recent years, the application of optical fiber communication technology is experiencing a period of rapid growth, which has created a demand for health monitoring techniques for fiber networks [1]. Optical reflectometry is one of the techniques developed for such a demand to detect positions of poor contacts and/or damages in the fiber by measuring the reflectivity distribution along the fiber under test (FUT). Compared to the conventional electrical sensors, optical reflectometry has numerous advantages including small size, long measurement range, capability of distributed measurement, electromagnetic resistance, and no need of power sources at sensing positions.

Widely known optical reflectometry techniques, such as optical time-domain reflectometry (OTDR) [2-6] and optical frequency-domain reflectometry (OFDR) [7-11], have the advantages of long measurement range and high spatial resolution, respectively. However, in OTDR, it is inevitable to repeat thousands of times of pulse injection and data averaging to obtain a high signal-to-noise ratio (SNR), which leads to a relatively low sampling rate. On the other hand, in OFDR, there exists a trade-off relation between the bandwidth of the light source and the measurement range, which puts forward high requirements for the light source. Although some solutions have been reported [10], the setup requires additional interferometry and becomes complicated and high in cost.

As an alternative of OTDR and OFDR, optical correlation-domain reflectometry (OCDR) has been developed on the basis of the synthesis of optical coherence functions (SOCF) [12], which can mitigate the drawbacks of conventional systems. OCDR can perform real-time distributed reflectivity measurement at a relatively low cost and has a unique advantage of random accessibility [12-23]. In a standard OCDR configuration, the frequency of the light reflected from the FUT (or the reference light) is shifted by tens of megahertz using an acousto-optic modulator (AOM) to perform optical heterodyne detection, which can reduce low-frequency noise disturbance [17]. To date, several approaches have been developed to simplify the standard OCDR for further reduction in implementation cost. For instance, an AOM-free configuration was developed by monitoring the foot of the reflected peak in the spectrum [24]. Another simplified configuration without the direct use of a reference path was also reported [25].

In the conventional AOM-free OCDR, an electrical spectrum analyzer (ESA) was used to process the electrical signals, which suffered from the following three problems. First, the ESA is generally bulky and high in cost, being an obstacle to system simplicity. Second, the zero-span function in the ESA is used to monitor the power of a particular frequency component of the foot of the reflected peak, where it is converted into a time-domain signal with a larger width due to the built-in low-pass filters. Therefore, the enhancement of spatial resolution of the OCDR will be limited. Lastly, data averaging for disturbance suppression is required for ESA-implemented OCDR configurations, which leads to a trade-off between the repetition rate and the spatial resolution.

In this work, to tackle these problems, we newly develop an ESA-free OCDR configuration. First, we numerically simulate the operations of the ESA-free and conventional AOM-free configurations and theoretically analyze their performances on distributed reflectivity measurement. Subsequently, we experimentally demonstrate the basic operation of the ESA-free OCDR and clarify that the spatial resolution can be improved by ~3 times, which is in good agreement with the simulation results. The results of both simulation and experiment are obtained at the repetition rate of 100 Hz, which equals the fastest record ever reported. Lastly, in a comparison experiment, we increased the repetition rate to up to 1 kHz to verify the effectiveness of the ESA-free OCDR on performing higher-speed measurement with unaffected spatial resolution.

## 2. Principles

The conceptual setup of the conventional AOM-free OCDR and the proposed ESA-free OCDR are shown in Fig. 1. The whole setup can be divided into an optical system (marked with blue solid lines) and an electrical system (marked with black solid lines). The two configurations share the same optical system, in which the optical output from a laser diode (LD) is divided into incident and reference light beams. The incident light is injected into the FUT, and the reflected signal is directed into a photodetector (PD), where the optical beat signal of the reference light and the reflected light is converted into an electrical signal. In the electrical system, the conventional OCDR and the ESA-free OCDR differ in the use of an ESA. In the conventional OCDR, the beat signal is first analyzed using an ESA, where the zero-span function is used to monitor the power of a particular frequency component of the foot of the reflected peak, which is subsequently displayed on an oscilloscope (OSC) to present the reflectivity of the sensing point.

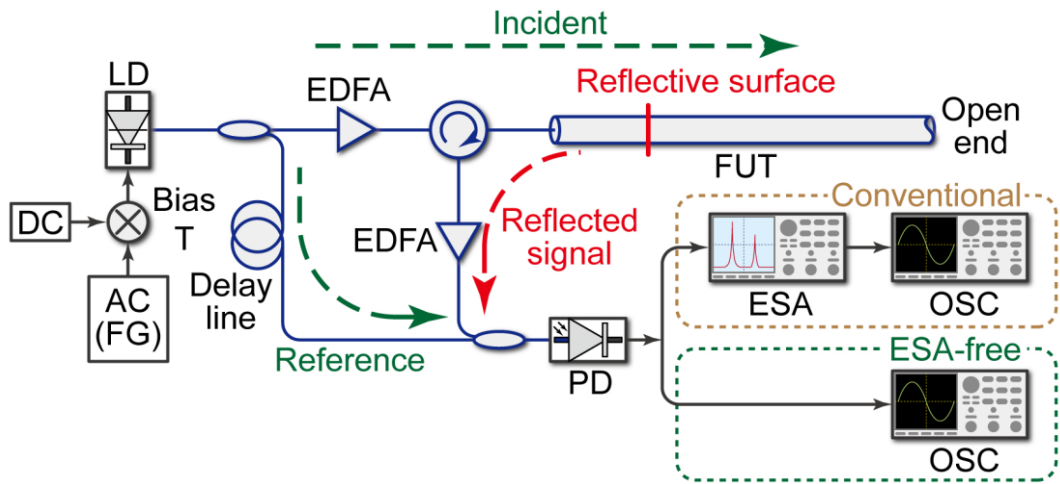

Fig. 1: Conceptual setup of the conventional and ESA-free OCDR.

To resolve the sensing position along the FUT, the optical frequency is often modulated in a sinusoidal waveform by directly modulating the injection current of the LD [15-18]. Then, the coherence function between the interfering two light beams, which is defined as a function of coherence degree versus optical path difference [16], is synthesized into a series of periodical delta-function-like correlation peaks [20]. We control the modulation frequency to generate only a single correlation peak, which can be regarded as a sensing point, within the range of the FUT to selectively observe the light reflected from the position of the correlation peak. By sweeping the modulation frequency, the position of the correlation peak is scanned along the FUT, and thus the reflectivity distribution can be obtained. The measurement range $D$ and the spatial resolution $\Delta z$ in a standard SOCF-OCDR system are given by [26,27]:

$$D = \frac{c}{2nf_m}, \tag{1}$$

$$\Delta z \cong \frac{0.76c}{\pi n \Delta f}, \tag{2}$$

respectively, where $c$ is the light velocity in vacuum, $n$ is the refractive index of the fiber core, $f_m$ is the modulation frequency, and $\Delta f$ is the modulation amplitude.

In the conventional OCDR, the built-in low-pass filters in the ESA can be used to adjust the shape of the monitored spectrum. However, the bandwidth of the spectrum is meanwhile broadened due to the effect of the low-pass filters, which becomes more significant as the measurement speed increases. Therefore, the use of the ESA deteriorates the spatial resolution and limits the performance at high-speed operation.

To mitigate the demerits of the conventional OCDR, we propose a further simplified ESA-free OCDR configuration, in which the electrical beat signal is directly transmitted to the OSC, skipping the ESA. In the OSC, the "peak detect" function is used to observe the intensity of the time-domain beat signal, which is given by

$$I_{\text{beat}}(t) = I_{\text{ref}}(t) + I_{\text{sig}}(t) + 2\sqrt{I_{\text{ref}}(t) \cdot I_{\text{sig}}(t)} \cdot J_0\left(2\frac{\Delta f}{f_m}\sin(\pi f_m \tau)\right), \tag{3}$$

where $I_{\text{ref}}(t)$ and $I_{\text{sig}}(t)$ are the intensities of the reference light and the signal light over time, respectively, $J_0$ is the Bessel function of the first kind of order zero, and $\tau$ is the delay of the transmission time between the reference and signal lights. In OCDR configurations with direct frequency modulation, the laser output inevitably suffers from additional amplitude modulation [28]. Therefore, $I_{\text{ref}}(t)$ and $I_{\text{sig}}(t)$ are given by

$$I_{\text{ref}}(t) = I_{\text{ref}} + \alpha \sin(2\pi f_{\text{m}} t), \tag{4}$$

and

$$I_{\text{sig}}(t) = I_{\text{sig}} + \alpha \sin(2\pi f_{\text{m}}(t+\tau)), \tag{5}$$

respectively, where $I_{\text{ref}}$ and $I_{\text{sig}}$ are the average intensity values of the reference light and the signal light, respectively, and $\alpha$ is the amplitude modulation coefficient. In short, the essential difference of the conventional and ESA-free configurations lies in the observed frequency band, where the conventional OCDR observes the power of the signal with a fixed frequency, while the ESA-free OCDR observes the full frequency band of the beat signal.

## 3. Simulation

To numerically analyze the feasibility of the ESA-free OCDR, we first performed a simulation on the operation of both the ESA-free and conventional OCDR configurations. The models of the setups used in the simulation are the same as those in Fig. 1. We adjusted the power of the reflected signal to the same level as the reference light power. The model of the FUT is shown in Fig. 2. Along the fiber, we assumed the reflectivity to be 0 except for the position of the reflective surface at 10 m. Distributed reflectivity measurement along the 0–35.8-m section of the FUT was performed. The number of sampling points was 2,500 at a sampling rate of 250 kHz. The laser was assumed to have an optical wavelength of 1,550 nm and a linewidth of 2 MHz. The refractive index of the core of the FUT was 1.5. The modulation amplitude was 1.9 GHz, which led to a theoretical spatial resolution of 2.6 cm in a standard OCDR according to Eq. (2). In the simulation of the conventional OCDR, the central frequency of the zero-span mode of the ESA was 1 MHz. The video bandwidth (VBW) of the ESA was 30 kHz.

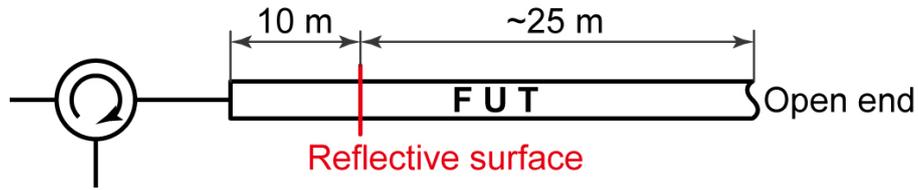

Fig. 2: Model of the FUT used in the numerical simulation.

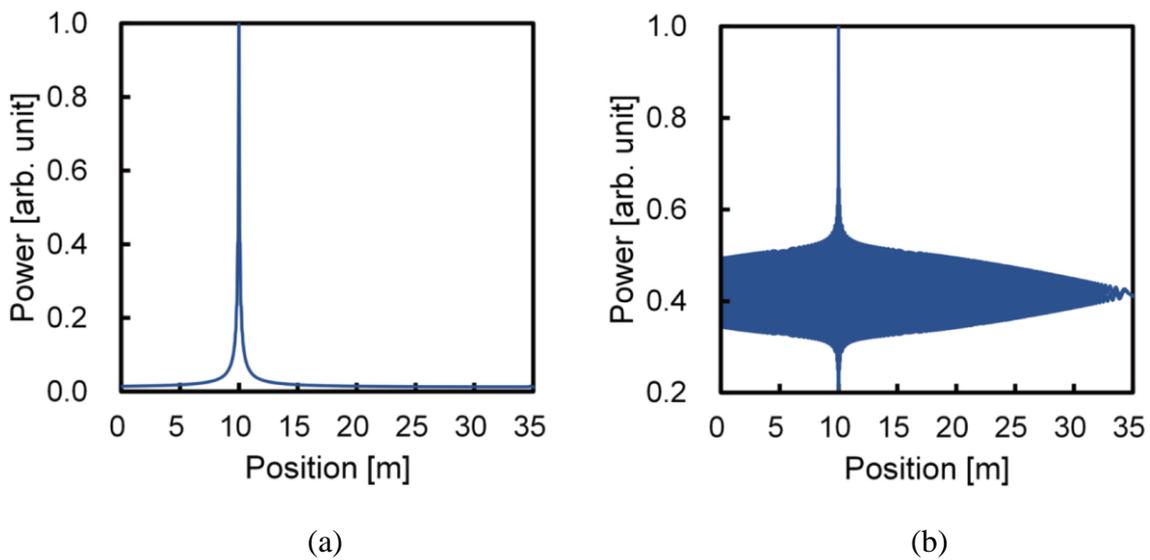

(a)  (b)

Fig. 3: Simulation results of normalized power distributions of the reflected signal in (a) conventional OCDR and (b) ESA-free OCDR.

In the conventional OCDR configuration, the power of the reflected signal was obtained by finding the power corresponding to the central frequency of the zero-span mode on the beat spectrum. On the other hand, in the ESA-free configuration, it was obtained by directly calculating the intensity of the beat signal using Eq. (3). The normalized power distributions of the reflected signal numerically obtained in the conventional and ESA-free configurations are shown in Fig. 3(a) and 3(b), respectively. The distribution curve obtained with the conventional OCDR was normalized by setting the power of the noise floor to 0 and the maximum power to 1, while the distribution curve obtained with the ESA-free OCDR was normalized by setting minimum and maximum power to 0 and 1, respectively. Both configurations effectively detected the reflective surface at 10 m. In the ESA-free OCDR, the reflection peak appeared in both positive and negative directions, with a noise floor sandwiched in the middle. Considering that the positive peak showed a higher SNR, we zoomed in on the positive side of the data as the result of actual observation.

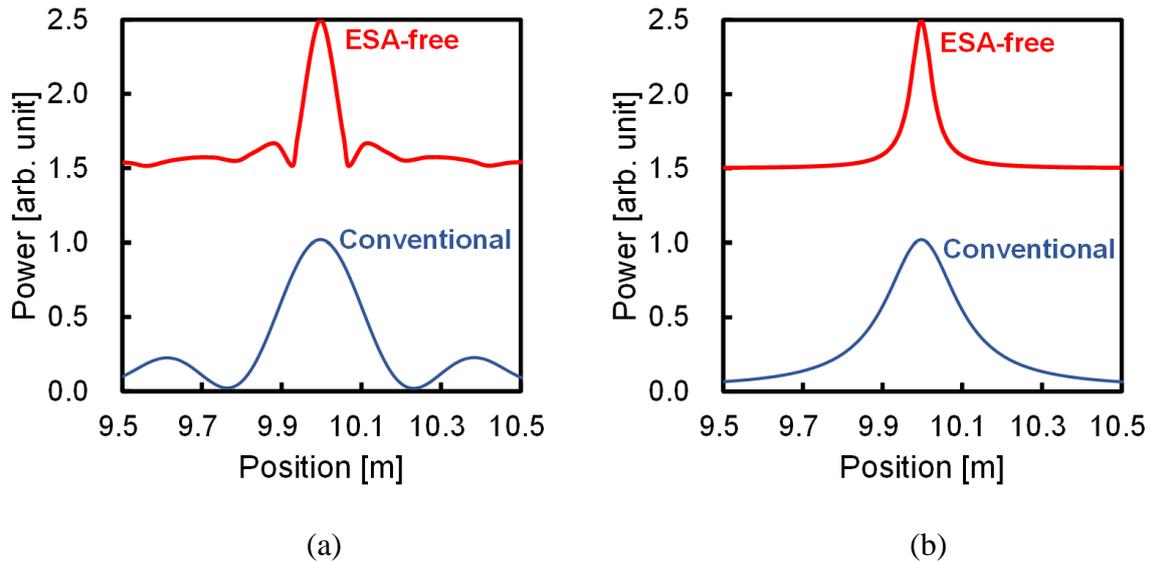

(a)                 (b)

Fig. 4: Simulation results of (a) normalized power distribution curves of the reflected signal and (b) Lorentzian fitted curves of the distribution curves near the reflective surface, obtained in both the conventional (blue) and ESA-free OCDR configurations (red; elevated by 1.5).

The normalized power distribution curves of the reflected signal near the reflective surface in the conventional and ESA-free configurations are shown in Fig. 4(a). For the conventional OCDR, the result had been filtered by the built-in low pass filter (VBW). In contrast, for the ESA-free OCDR, the result was obtained by applying the peak-detect function of the OSC on the data near the positive reflection peak and then normalized by setting the maximum power to 1 and the power of the noise floor to 0. Subsequently, the power distribution curve of the ESA-free OCDR was elevated by 1.5 overall to be clearly distinguished from that of the conventional OCDR. To quantitatively compare the spatial resolutions of the two configurations, we fitted the distribution curves using Lorentzian functions, as shown in Fig. 4(b). The full width at half maximum (FWHM) of the conventional and ESA-free OCDR are 21.7 cm and 6.2 cm, respectively, which shows the potential of the ESA-free OCDR to improve the spatial resolution by 3.5 times compared to the conventional one. The significant discrepancy of the spatial resolution between the obtained value of 6.2 cm and the theoretical value of 2.6 cm can be possibly attributed to the direct modulation scheme, where intensity modulation is inevitably induced and deteriorates the measurement performance.

## 4. Experiments

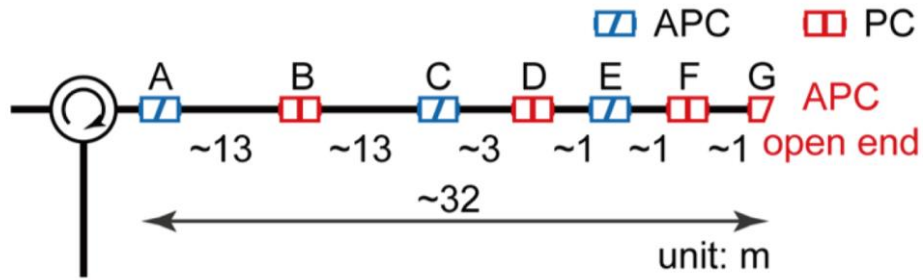

Fig. 5: Structure of the FUT used in the experiment.

To verify the theoretical analysis of the ESA-free OCDR, we experimentally implemented both the ESA-free and conventional OCDR systems to perform identical distributed reflectivity measurements. The experimental setups of the two configurations are the same with that in Fig. 1 used in the simulation, except for an additional polarization controller on the reference path of the experimental configuration. The structure of the FUT is shown in Fig. 5. The optical frequency of the laser output was 193.77 THz, and the laser linewidth was 2 MHz. The length of the delay line was 3 km. The modulation frequency was swept from 2.017 MHz to 2.119 MHz to perform a distributed sensing along a section of approximately 32 m on the FUT. The sweeping period of the modulation frequency was 10 ms (repetition rate: 100 Hz), which is the same as the fastest record previously reported [22]. The FUT was designed using fiber segments with the lengths of 13 m, 3 m, and 1 m, which were connected with physical contact (PC) connectors and angled physical contact (APC) connectors. The reflection attenuation of the PC connectors was >25 dB, higher than that of the APC connectors of >60 dB. In the ESA-free configuration, the peak-detect function of the OSC was used to monitor the power of the reflected peak of each sampling point. The refractive index of the core of the FUT was 1.5. The modulation amplitude was 1.9 GHz. In this case, the theoretical spatial resolution in a standard OCDR would be 2.6 cm according to Eq. (2). In the conventional configuration, the central frequency of the zero-span function was set to 1 MHz, and the resolution bandwidth and the VBW of the ESA were set to 300 kHz and 30 kHz, respectively.

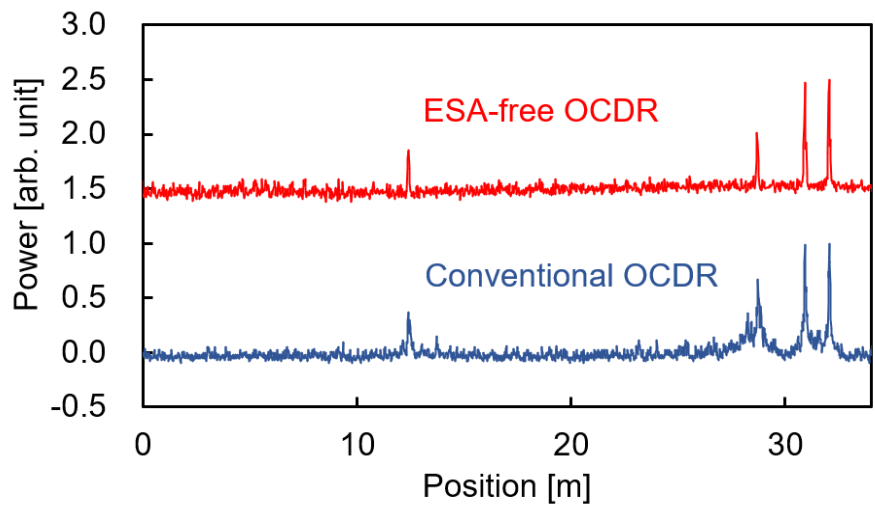

Fig. 6: Normalized reflected power distributions along the whole FUT measured using both conventional OCDR (blue) and ESA-free OCDR (red; elevated by 1.5).

The measured reflectivity distributions by the two configurations are shown in Fig. 6. The horizontal axis shows the relative position along the fiber, the zero point of which was set to the position of Connector A. The power distributions of the reflected peak were normalized by setting the noise floor to 0 and the maximum power to 1, and the power distribution of the ESA-free OCDR was elevated by 1.5 overall. Both configurations effectively detected the points with high reflectivity at Connector B, D, and F, and the open end of the FUT. In addition, the widths of the reflected peaks on the distribution curve of the conventional OCDR was wider than those of the ESA-free configuration.

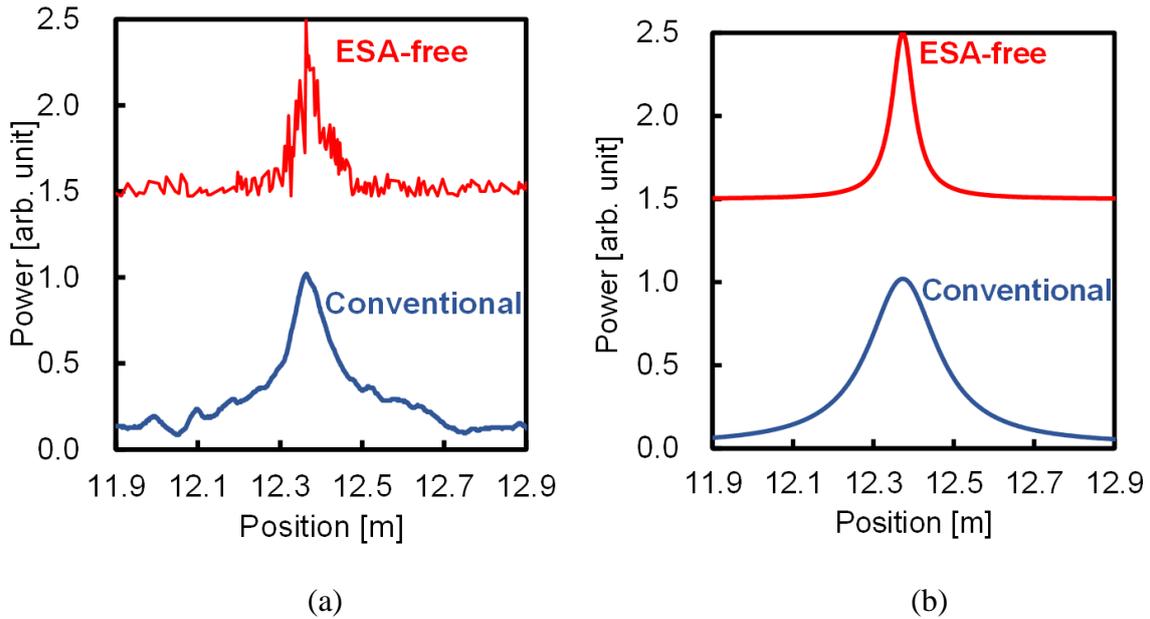

Fig. 7: Experimental results of (a) normalized power distribution curves of the reflected signal and (b) Lorentzian fitted curves of the distribution curves near the reflective surface, obtained in both the conventional OCDR (blue) and ESA-free OCDR (red; elevated by 1.5).

To quantitatively compare the performances of the two configurations, we zoomed in on the experimental results near Connector B, which is sufficiently far from other connectors and not influenced by other reflected peaks. In the conventional OCDR, data was averaged 16 times, while in the ESA-free OCDR, no data averaging was performed. The power distributions near Connector B are shown in Fig. 7(a). It was clear that the width of the reflected peak was narrower in the ESA-free configuration. Subsequently, we fitted the power distribution curves near the reflected peak using Lorentzian functions. The fitting results are shown in Fig. 7(b). The FWHM of the fitted curves of the conventional and ESA-free OCDR were 22.9 cm and 7.0 cm, respectively. The experimental results are highly consistent with the simulation results and show that the spatial resolution of the ESA-free OCDR can be improved by ~3 times even without the need of data averaging.

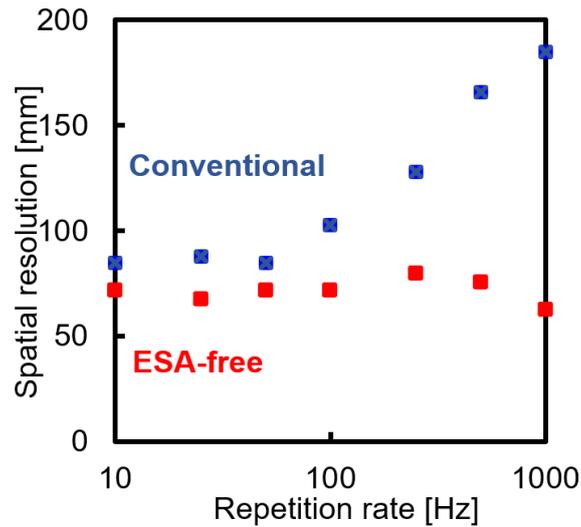

Fig. 8: Measured FWHM values of the reflected peaks obtained at repetition rates varying from 10 Hz to 1 kHz in both conventional OCDR (blue) and ESA-free OCDR (red).

Lastly, considering that data averaging for disturbance suppression is not required in the ESA-free OCDR, we further increased the repetition rate to verify the effectiveness of the ESA-free OCDR on performing high-speed distributed reflectivity measurement with a higher spatial resolution. We experimentally compared the FWHM values of the reflected peaks measured at repetition rates varying from 10 Hz to 1 kHz in both conventional and ESA-free OCDR configurations. The experimental setup is the same as the former one, as shown in Fig. 5. Except that the VBW of the ESA was 300 kHz in the conventional OCDR configuration, all the other experimental conditions remained unchanged. The measured FWHM values of reflected peaks in both ESA-free and conventional OCDR are shown in Fig. 8. The FWHM measured with the conventional OCDR at 100 Hz was 10.3 cm, which was largely improved compared to that in the former experiment. The performance of the conventional OCDR was improved because of the tuning of the polarization controller and the observation of the high-frequency component of the signal with increased VBW. Figure 8 shows that, when the repetition rate was lower than ~100 Hz, the spatial resolution was slightly improved in the ESA-free OCDR than that in the conventional one, and the resolutions in both cases were almost constant regardless of the repetition rate. However, with the increase of the repetition rate over 100 Hz, the FWHM of the conventional OCDR was broadened significantly, while that of the ESA-free OCDR remained almost the same. At the repetition rate of 1 kHz, the FWHM values of the ESA-free and conventional OCDR configurations were 6.3 cm and 18.5 cm, respectively. Thus, we showed that the trade-off relationship between the repetition rate and the spatial resolution can be mitigated using the ESA-free OCDR.

## 5. Conclusions

We proposed a high-speed ESA-free OCDR and demonstrated by both simulation and experiments that, in addition to the system simplification, the spatial resolution can be improved because of the elimination of the effect of the built-in filters in the ESA. Simulation results and experimental results showed a good agreement, which indicated that the ESA-free configuration can improve the spatial resolution by approximately 3 times compared to that of the conventional OCDR. Lastly, we experimentally verified the effectiveness of the ESA-free OCDR at higher repetition rates and achieved a spatial resolution of 6.3 cm at a repetition rate of as high as 1 kHz, which indicates that the trade-off relationship between the measurement speed and the spatial resolution in conventional systems can be significantly mitigated by the ESA-free OCDR. Note that the repetition rate achieved experimentally in this work was limited by the performance of the function generator. By replacing it with a voltage-controlled oscillator, even higher-speed operation may be feasible.

**Declaration of Competing Interest**

The authors declare that they have no known competing financial interests or personal relationships that could have appeared to influence the work reported in this paper.

**Acknowledgements**

This work was partially supported by the Japan Society for the Promotion of Science (JSPS) KAKENHI (Grant Nos. 21H04555, 22K14272, and 20J22160) and the research grants from the Murata Science Foundation, the Telecommunications Advancement Foundation, the Takahashi Industrial and Economic Research Foundation, the Yazaki Memorial Foundation for Science and Technology, and the Konica Minolta Science and Technology Foundation.

## Biographies

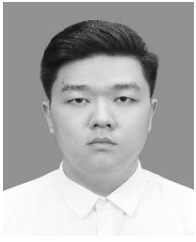

**Guangtao Zhu** received the B.E. degree in Light Source and Lighting from Taiyuan University of Technology, China, in 2020. Since 2020, he has been studying distributed fiber-optic sensing techniques for his M.E. degree in applied physics at Yokohama National University, Japan. He is a student member of the Japanese Society of Applied Physics (JSAP).

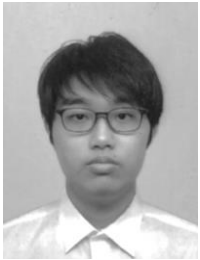

**Tomoya Miyamae** received the B.E. degree in electrical and computer engineering from Yokohama National University, Japan, in 2021.Since 2020, he has been studying distributed fiber-optic sensing techniques for his M.E. degree in electrical and computer engineering at Yokohama National University. He is a student member of the Japanese Society of Applied Physics (JSAP).

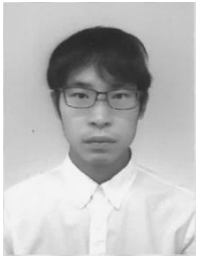

**Kohei Noda** received the B.E. and M.E. degrees in electrical and electronic engineering from Tokyo Institute of Technology, Japan, in 2018 and 2020, respectively. Since 2020, he has been studying distributed fiber-optic sensing techniques for his Dr.Eng. degree in electrical and electronic engineering at Tokyo Institute of Technology. He is a student member of the Japanese Society of Applied Physics (JSAP).

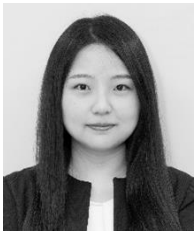

**Heeyoung Lee** received the B.E. degree in mechanical engineering from Kyungpook National University, Korea, in 2014, and the M.E. degree in information processing and Dr.Eng. degree in electrical and electronic engineering from Tokyo Institute of Technology, Japan, in 2017 and 2019, respectively. Since 2019, she has been an Assistant Professor at the College of Engineering, Shibaura Institute of Technology, Japan.

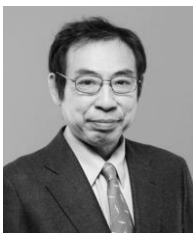

**Kentaro Nakamura** received the B.E., M.E., and Dr.Eng. degrees from Tokyo Institute of Technology, Japan, in 1987, 1989, and 1992, respectively. Since 2010, he has been a Professor at the Precision and Intelligence Laboratory (presently, Institute of Innovative Research), Tokyo Institute of Technology. His research field is the applications of ultrasonic waves, measurement of vibration and sound using optical methods, and fiber-optic sensing.

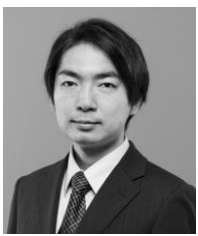

**Yosuke Mizuno** received the B.E., M.E., and Dr.Eng. degrees in electronic engineering from the University of Tokyo, Japan, in 2005, 2007, and 2010, respectively. From 2012 to 2020, he was an Assistant Professor at Tokyo Institute of Technology. Since 2020, he has been an Associate Professor at the Faculty of Engineering, Yokohama National University, Japan, where he is active in fiber-optic sensing and polymer optics.